# Thermoelectric behavior of Ruddlesden-Popper series iridates

I. Pallecchi [1], M. T. Buscaglia [2], V. Buscaglia [2], E. Gilioli [3], G. Lamura [1], F. Telesio [1], M. R. Cimberle [1], D. Marré [1]

[1] CNR-SPIN and Dipartimento di Fisica, Via Dodecaneso 33, 16146 Genoa, Italy
[2] CNR-IENI, Via De Marini 6, 16149 Genoa, Italy
[3] CNR-IMEM, Area delle Scienze, 43124 Parma, Italy

**Abstract**
The goal of this work is studying the evolution of thermoelectric transport across the members of the Ruddlesden-Popper series iridates $Sr_{n+1}Ir_nO_{3n+1}$, where a metal-insulator transition driven by bandwidth change occurs, from the strongly insulating $Sr_2IrO_4$ to the metallic non Fermi liquid behavior of $SrIrO_3$. $Sr_2IrO_4$ (n=1), $Sr_3Ir_2O_7$ (n=2) and $SrIrO_3$ (n=∞) polycrystals are synthesized at high pressure and characterized by structural, magnetic, electric and thermoelectric transport analyses. We find a complex thermoelectric phenomenology in the three compounds. Thermal diffusion of charge carriers accounts for the Seebeck behavior of $Sr_2IrO_4$, whereas additional drag mechanisms come into play in determining the Seebeck temperature dependence of $Sr_3Ir_2O_7$ and $SrIrO_3$. These findings reveal close relationship between magnetic, electronic and thermoelectric properties, strong coupling of charge carriers with phonons and spin fluctuations as well as relevance of multiband description in these compounds.

## 1. Introduction

Transition metal oxides exhibit an endless variety of physical properties, not yet fully explored, originating from the competition between comparable energy scales, namely the Coulomb interaction U, the amplitude of the hopping integral, the crystal field splitting and the spin-orbit coupling. The hierarchy of such energy scales rules the interaction between charge, spin and lattice degrees of freedom, thus determining the ground state of each system.

In iridium oxides, the Coulomb interaction U~0.5eV [1,2] is limited due to the extended character of *5d* orbitals, the octahedral crystalline field, which splits the five *d* orbitals into two higher $e_g$ and three lower $t_{2g}$ states, is as large as 3-4 eV [3], and the spin-orbit coupling ~0.4eV [4,5] is huge, as compared to typical values of tens of meV in *3d* transition metal oxides, and thus comparable with U. The magnitude of the spin-orbit coupling is the key to understand the electron band scenario. Indeed, since the seminal works by Kim [1,6] and Moon [2], the electronic structure has been identified as described by the effective total angular momentum $J_{eff}$ states, whose effective quantum spin number $J_{eff}$ incorporates also the orbital momentum. In particular, the $t_{2g}$ states, hosting five electrons per formula unit, are split by the strong spin orbit interaction into a totally filled $J_{eff}=3/2$ quartet band and a half-filled $J_{eff}=1/2$ doublet band. The latter band is made up of an equal mixture of $d_{xz}$, $d_{yz}$ and $d_{xy}$ orbitals and has isotropic orbital and mixed spin character. It is pretty narrow [2], because the extension of the *d* shells points towards a strong coupling between the *d* orbitals and the neighboring oxygen orbitals, implying a tendency to form distorted structures with smaller Ir-O-Ir bond angle instead of the ideal ones, which yields a narrowing of the *d* bandwidth W. In this situation, even a small U opens a Mott gap, making iridium oxides $J_{eff}=1/2$ Mott insulators. The $J_{eff}=1/2$ Mott ground state has been experimentally identified by angle-resolved photoemission spectroscopy [7], resonant X-rays scattering [6], optical conductivity [1,2] and also confirmed by *ab initio* calculations [2,5,8]. This $J_{eff}=1/2$ Mott scenario explains why in the Ruddlesden-Popper series $Sr_{n+1}Ir_nO_{3n+1}$ insulating behavior is systematically observed, except for the limit perovskite compound n=∞ $SrIrO_3$, despite transport occurs in *5d* orbitals, whose extended character should favor large bandwidths and hybridization with *2p* orbitals, as compared to *3d* and *4d* orbitals.

More specifically, in the Ruddlesden-Popper series $Sr_{n+1}Ir_nO_{3n+1}$ the bandwidth W is changed with the series index n, representing the number of $SrIrO_3$ perovskite layers sandwiched between extra SrO layers. Indeed, W is proportional to the Ir coordination number z, i.e. number of neighboring Ir ions, whose value is z=4 in $Sr_2IrO_4$ (n=1), z=5 in $Sr_3Ir_2O_7$ (n=2) and z=6 in $SrIrO_3$ (n=∞). More rigorously, the behavior of the increasing bandwidth with the series index n has been calculated theoretically for $J_{eff}$=1/2 orbitals [8,9]. The bandwidth values measured by optical conductivity have turned out to be W=0.48 eV (n=1), 0.56 eV (n=2) and 1.01 eV (n=∞) and the transition from Mott insulator (W<<U) to correlated metal (W≥U) has been directly observed for compounds with n between 2 and ∞ [2]. Intriguingly, contrasting experimental evidence has been found by angle-resolved photoemission [10], which indicates that the bandwidth of $SrIrO_3$ is instead narrower than that of its insulating two-dimensional counterpart $Sr_2IrO_4$, as a consequence of the combination strong spin-orbit interactions, dimensionality, and both in- and out-of-plane $IrO_6$ octahedral rotations. Whereas the ratio U/W tunes the effects of correlation in the Ruddlesden-Popper series, also the dimensional character varies with increasing Ruddlesden-Popper series index n. The dimensionality increases from $Sr_2IrO_4$ (n=1), which is a two-dimensional (2D) Mott-like antiferromagnetic insulator, to $SrIrO_3$ (n=∞), which is a three dimensional (3D) correlated metal. The $Sr_3Ir_2O_7$ (n=2) compound, which is a narrow gap antiferromagnetic insulator, is on the verge of a dimensional crossover, as well as on the verge of a metal-insulator Mott transition.

In this work, we plan to get information about the competition of the relevant energy scales from the analysis of macroscopic magnetic, electronic, thermoelectric and thermal transport properties and their intimate inter-relationship. Our experimental study is carried out on polycrystalline samples of the Ruddlesden-Popper series $Sr_{n+1}Ir_nO_{3n+1}$ (n=1, 2 and ∞).

## 2. Experimental

Polycrystalline $Sr_{n+1}Ir_nO_{3n+1}$ (n=1, 2 and ∞) samples are synthesized from SrO and $IrO_2$ powders (Alfa Aesar) under high pressure/high temperature (HP/HT) conditions in a in a 6/8-type multi anvil apparatus (Rockland Research Corporation), at temperature 1000 °C and pressure 50 kbar. The different samples are individually encapsulated in Au using 50 μm thick foils and stacked in a 150 $mm^3$ cell assembly to guarantee the same synthesis conditions. The pressure is firstly increased at a rate of 120 bar/min; then the capsule is heated at 50 ° C/min, kept for 1 hour and quenched to room temperature by switching off the heater. The pressure is finally slowly released at -0.4 bar/min. The 1 mm thick and 5 mm diameter disk-shaped samples are recovered by removing the Au capsule with a blade.

The phase purity of final samples is checked by X-ray diffraction (XRD, CoKα radiation, ips PW1710, Philips, Eindhoven, The Netherlands). The microstructure of the samples is investigated by scanning electron microscopy (SEM) using a LEO 1450VP microscope (LEO Electron Microscopy Ltd., Cambridge, U.K.) equipped with an Energy Dispersive X-ray Analyzer (EDS) INCA Energy 300 (Oxford Instruments Analytical, U.K.) for elementary composition measurements. Magnetization curves as a function of magnetic field up to 55000 Oe and temperatures from 5K to 330K are measured in a SQUID magnetometer by Quantum Design. For transport measurements, the disk shaped pellets are cut into parallelepiped bars of around 1mm X 1mm cross section and 5 mm length, by means of a diamond saw. Resistivity, magnetoresistivity, Hall effect, Seebeck effect and thermal conductivity data are measured in a PPMS (Physical Properties Measurement System) by Quantum Design, equipped with thermal transport option, at temperatures from room temperature down to 5K and in magnetic fields up to 70000 Oe. More specifically, resistivity is measured using a standard four probe method, Hall coefficient ($R_H$) is determined by measuring the transverse resistivity at selected fixed temperatures, sweeping the field from -70000 Oe to 70000 Oe, and Seebeck (S) effect and thermal conductivity are measured in the continuous scanning mode with a 0.4 K/min cooling rate. Regarding Hall effect measurements, due to the unfavorable sample geometry (1 mm thick bars), we estimate an experimental uncertainty on carrier density data points up to 50%, especially for the largest values of carrier density. Geometrical issues related to the polycrystalline nature of the samples, namely random orientation of crystallites and porosity, may affect the extraction of resistivity absolute values, considering resistivity anisotropy values up to 7 measured in $Sr_3Ir_2O_7$ single crystals [11] and between 3 and 5.6 measured in $Sr_2IrO_4$ single crystals [12].

## 3. Results and discussion
### 3.1 Structural, morphological and magnetic characterization

In figure 1, we present X-rays diffraction patterns collected on our $Sr_2IrO_4$ (n=1), $Sr_3Ir_2O_7$ (n=2) and $SrIrO_3$ (n=∞) polycrystals, where the most intense Bragg reflections of the main phases are labelled with corresponding Miller indices. For the $Sr_2IrO_4$ sample, the main phase with tetragonal symmetry (space group $I4_1/acdz$) is clearly dominant, with small additional peaks of $SrCO_3$, Ir and $Sr_3Ir_2O_7$ secondary phases. The contribution of this latter secondary phase is detected also in magnetic measurements (see figure 3a)), but not in transport properties. Indeed, SEM/EDS observation show that this secondary phase appears in separated grains embedded in the matrix of the main $Sr_2IrO_4$ phase, so that these grains do not form a low resistance percolative path for transport along the sample. Therefore, all the electric and thermoelectric characterization of this sample presented hereafter is representative of the $Sr_2IrO_4$ phase. For the $Sr_3Ir_2O_7$ sample, the main phase with tetragonal structure is largely dominant. For this sample, small secondary peaks of $Sr_4Ir_3O_{10}$, $SrIrO_3$ and Ir are present, whose possible minor contribution to magnetic and transport properties is discussed in the following sections. Finally, for the $SrIrO_3$ sample, the main orthorhombic perovskite phase (space group Pnma) is again by far dominant, with very small additional peaks of $SrCO_3$, $IrO_2$ and Ir secondary phases, giving negligible magnetic and transport contributions to the measured properties. Indeed, it is known that $SrIrO_3$, if prepared at ambient pressure, crystallizes in a hexagonal structure with monoclinic distortion [13], however it turns out orthorhombic if prepared by high pressure synthesis [14] or if deposited in the form of epitaxial film on suitable substrates [15]. As the index *n* of the Ruddlesden-Popper series $Sr_{n+1}Ir_nO_{3n+1}$ represents the number of $SrIrO_3$ perovskite layers sandwiched between extra SrO layers, the orthorhombic $SrIrO_3$ phase represents the end member of the series, having no extra SrO layers. We point out that most polycrystalline samples of literature inevitably contain small amounts of secondary phases [14,16,17,18,19,20]. In our samples, the total amount of secondary phases is estimated to be few percent. For all the samples, the EDS analyses confirm the XRD results, indicating Sr/Ir ratios in fair agreement with the composition of the expected phase in each case, namely the measured average ratios result <Sr/Ir>=2.03, <Sr/Ir>=1.48 and <Sr/Ir>=1.06 for $Sr_2IrO_4$, $Sr_3Ir_2O_7$ and $SrIrO_3$, respectively.

In figure 2, SEM images of the microstructures of the $Sr_2IrO_4$ (n=1), $Sr_3Ir_2O_7$ (n=2) and $SrIrO_3$ (n=∞) polycrystals are shown. The phases with n = 1 and n = 2 are composed of rectangular tabular crystals typical of the Ruddlesden-Popper phases, whereas the $SrIrO_3$ sample shows equiaxed grains with predominant cubic morphology typical of perovskites. From the images it is also apparent that our samples approach full density, thanks to the high pressure synthesis.

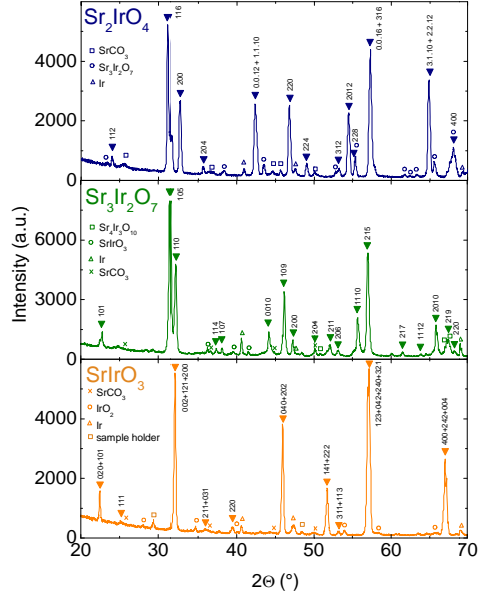

**Figure 1:** X-rays diffraction patterns measured in the $Sr_2IrO_4$ (n=1), $Sr_3Ir_2O_7$ (n=2) and $SrIrO_3$ (n=∞) polycrystals. The filled triangles (▼) indicate Bragg reflections of the main phases and corresponding indices.

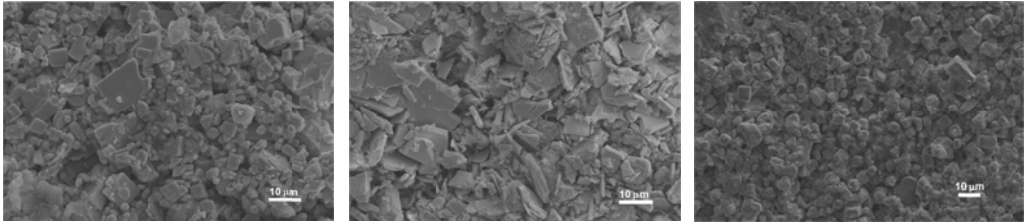

**Figure 2:** From left to right, SEM images of the microstructure of the $Sr_2IrO_4$ (n=1), $Sr_3Ir_2O_7$ (n=2) and $SrIrO_3$ (n=∞) polycrystals.

The measured magnetic data of our $Sr_2IrO_4$, $Sr_3Ir_2O_7$ and $SrIrO_3$ polycrystals (shown in the left panels of figure 3 and more extensively presented and discussed in the supplementary data) are consistent with existing literature. Specifically, the magnetic moment curves m(T) and m(H) of the $Sr_2IrO_4$ n=1 member of the series are consistent with the expected weak ferromagnetism associated to canted AFM at temperatures below $T_N≈200K$ [6,12], while those of the $Sr_3Ir_2O_7$ n=2 member of the series confirms the existence of AFM canted ordering below $T_N≈290K$ [8,21,22]. For both $Sr_2IrO_4$ and $Sr_3Ir_2O_7$, the extracted ordered magnetic moments per Ir atom turn out to be a small fraction of the value predicted for the $J_{eff}=1/2$ state, as observed in literature [11,12,16,23], which is a consequence of AFM exchange interactions and structural details [24]. The m(T) curves of $Sr_3Ir_2O_7$ show an additional feature around T*≈240K. Moreover, we observe that in a temperature range around T*, the hysteretic m(H) loops exhibit an exchange bias effect and a negative contribution to magnetoresistivity appears. These anomalies may form as a consequence of a magnetostrictive distortion manifesting just around T* [11] and need further investigations. The $SrIrO_3$ end member of the series exhibits paramagnetic behavior well described by a modified Curie-Weiss law below ~200K (at 200K M/H curves present an inflection point, similarly observed at ~170K in ref. [14]), with an effective magnetic moment per Ir atom $\mu_{eff}=0.0025\mu_B$ and a Curie temperature ≈1K, close to zero. Such small Curie temperature suggests that the magnetic interaction between the two electronic spins on the nearest-neighbor Ir sites lies at the crossover between FM-type and AFM-type. S-shaped m(H) curve of $SrIrO_3$ measured at low temperature, shown in the left inset of panel c) in figure 3, indicates exchange enhanced paramagnetism [14].

## 3.2 Electric transport characterization

Before introducing thermoelectric properties, it is necessary to gather information not only about magnetic behaviors, but also about electric transport, as the microscopic mechanisms determining these properties are closely interdependent. In this section, electric transport properties of the Ruddlesden-Popper series iridates are presented.

In figure 3d), the resistivity of the lowest index (n=1) member of the series $Sr_2IrO_4$ is plotted in logarithmic scale. It exhibits insulating behavior, with values that span six orders of magnitude, ranging from $\sim 10^{-3}$ $\Omega$m at room temeprature to $\sim 10^3$ $\Omega$m at low temperature. As expected, this is a Mott insulator, and thus the less conducting sample of the series. The shape and magnitude of resistivity curves and highly non linear current-voltage characteristics (shown in supplementary data), in substantial agreement with literature (see for example the stoichiometric crystal of ref. [25]), rule out any significant influence of iridate secondary phases on transport. However, when comparing with the scattered literature data, non-linearity [12] and sensitivity to oxygen stoichiometry [25], as well as unknown geometrical factors in polycrystals, must be taken into account. By fitting the resistivity temperature dependence, we identify different regimes, similarly to ref. [12,23,26]. Two distinct variable range hopping (VRH) regimes with slightly different parameters in the temperature ranges 5K-40K and 40K-120K are recognized as linear trends in the plots displayed in the insets of figure 3d). Indeed, in the VRH model, the resistivity $\rho$ is predicted to follow the law $\rho \propto \exp\left(\frac{T_0}{T}\right)^{\frac{1}{D+1}}$, where $T_0$ is a constant and D is the dimensionality of the system. The two insets refer to either 2D or 3D VRH models, which provide comparable fit quality, hence it cannot be concluded whether 2D or 3D VRH better fits the experiment. Between 140K and 310 K a logarithmic behavior $\ln(\rho) \propto -T$ is identified as a linear trend in the third inset of figure 3d), as in ref. [26], but its physical origin is unclear. From measurements of current-voltage characteristics (shown in supplementary data) it turns out that departure from ohmic behavior increases with decreasing temperature and is dramatic at the lowest temperatures below 50K, where negative differential resistivity, not explained by Joule heating effects alone, is observed, as in ref. [12]. In this respect, we point out that all the transport measurements are performed in the low electric field ohmic regime.

Figure 3e) reports transport characterization of $Sr_3Ir_2O_7$, the crossover compound among n=1, n=2 and n=∞ members of the series, that is the closest to the metal-insulator transition on the insulating side. The resistivity $\rho$ exhibits semiconducting behavior, but with a much milder temperature dependence as compared to the $Sr_2IrO_4$ compound. Indeed, the $\rho$ values span little more than one order of magnitude between room temperature and low temperature in the range $10^{-3}$-$10^{-4}$ $\Omega$m. This variation is sizably smaller than in most literature data, where it is in general 3-4 orders of magnitude [11,16,27,28]. For this discrepancy, there may be reasons related to oxygen stoichiometry, as for the $Sr_2IrO_4$ compound, random orientation of crystallites (for example the resistivity variation along the c axis is smaller than along the ab planes [11,27] and possibly it may be also current dependent as in the $Sr_2IrO_4$ compound [25]), but also a contribution of the $SrIrO_3$ impurity phase. Clear changes of slope are observed around 240K and 290K, in correspondence of the magnetic transitions at T* and $T_N$, in agreement with what is widely reported in literature [11,16,22,27,29]. The current-voltage characteristics (shown in supplementary data) are ohmic at high temperature, while mild non-linearity appears only below 30K, consistent with ref. [29]. As pointed out above, some amount of $SrIrO_3$ impurity phase could play a role in decreasing the measured resistivity. However, from the observation of all the transport data shown in the main paper and in supplementary data, it turns out that the overall shapes of transport curves must be representative of the correct $Sr_3Ir_2O_7$ main phase, in that resistivity shows semiconducting behavior (as opposed to metallic behavior of the $SrIrO_3$ phase), current-voltage characteristics are non linear at low temperature as in literature [29] (as opposed to ohmic ones of the $SrIrO_3$ phase), magnetoresistance has a negative contribution in a temperature range around the magnetic transitions of the $Sr_3Ir_2O_7$ phase as in literature [29] (as opposed to positive cyclotron magnetoresistance of the SrIrO$_3$ phase) and changes of slope in the resistivity curve across the magnetic transitions of the Sr$_3$Ir$_2$O$_7$ phase are observed. Any percolating path of the SrIrO$_3$ impurity phase would electrically short the sample and mask these features. Finally, the tranport properties of the end member of the series SrIrO$_3$ are presented in figure 3f), where the metallic resistivity is shown, together with a power law fit. The law $\rho = \rho_0 + \rho_1 T^\alpha$, with $\rho_0$, $\rho_1$, and $\alpha<2$ constant parameters, characterizes a non-Fermi liquid behavior, where the Landau Fermi-liquid theory, describing the electrons in a metal at low energies as a collection of weakly interacting particles, is violated. This may occur as a consequence of the coupling with spin fluctuations. Our fit yields $\alpha\approx1$ and indeed for ferromagnetic spin fluctuations in 3D, $\alpha=1$ is predicted [30]. The observed positive downward curvature of the low temperature m(H) curves (shown in the left inset of figure 3c), similar to ref. [14], consistently points to enhanced paramagnetism. Below 50K, the resistivity upturn indicates mild charge localization. Indeed, in a single band free electron picture, we can calculate the Fermi wavevector $k_F$ and electron mean free path $l$ and it turns out that the product $k_F \cdot l$ is close to unity and $l$ is close to the unit cell paramter in this compound, which sets the Ioffe-Regel criterion for metallic behavior [31]. It must be cautiously pointed out that the polycrystalline nature of our sample may yield an overestimation of the resistivity, indeed resistivity values measured in films with similar $\rho(T)$ dependence are up to one order of magnitude smaller than our values [10,32,33,34,35] so that the real mean free path $l$ could be larger than the one extracted from the present data. Both non-Fermi liquid metallic behavior and low temperature upturn are consistent with literature data on SrIrO$_3$ films [32,36]. Current-voltage characteristics (shown in supplementary data) are ohmic at all temperatures for this compound, as expected from a metal.

In Figure 4, we report Hall resistance $R_H$ curves measured in the three compounds. The charge carrier density could extracted from the Hall resistance $R_H$ in a single band framework as $n = (qR_H)^{-1}$, however the single band description is inadequate for these compounds, and $R_H$ does at most indicate the type (either holes or electrons) of the dominant charge carriers and a rough idea of the temperature dependence of its concentration. For all the compounds the negative sign of the Hall resistance $R_H$ indicates that the dominant charge carriers are n-type and can be identified with $J_{eff}=1/2$ electrons of the upper Hubbard band. For Sr$_2$IrO$_4$, we find that the carrier density decreases significantly with decreasing temperature from $\sim 10^{19}$ cm$^{-3}$ at room temperature down to $\sim 10^{13}$ cm$^{-3}$ at the lowest temperature, consistently with the strongly insulating behavior of this compound. For Sr$_3$Ir$_2$O$_7$, the carrier density is weakly temperature dependent and around few times $10^{19}$ cm$^{-3}$, indicating proximity of Sr$_3$Ir$_2$O$_7$ to the metal-insulator transition. Finally, for SrIrO$_3$, the charge carrier density is around $10^{20}$ cm$^{-3}$ at low temperature, in agreement with literature [32].

We point out that a single band analysis of our magnetotransport data does not provides fully consistent results, in particular the temperature dependences of Hall mobility and of the mobility extracted from the cyclotron magnetoresistivity coefficient are not in agreement with each other (see supplementary data file). A more complex multiband framework based on several parameters, including density and mobility of each type of carriers, could account for the whole electric and thermoelectric (shown in the next section) phenomenology.

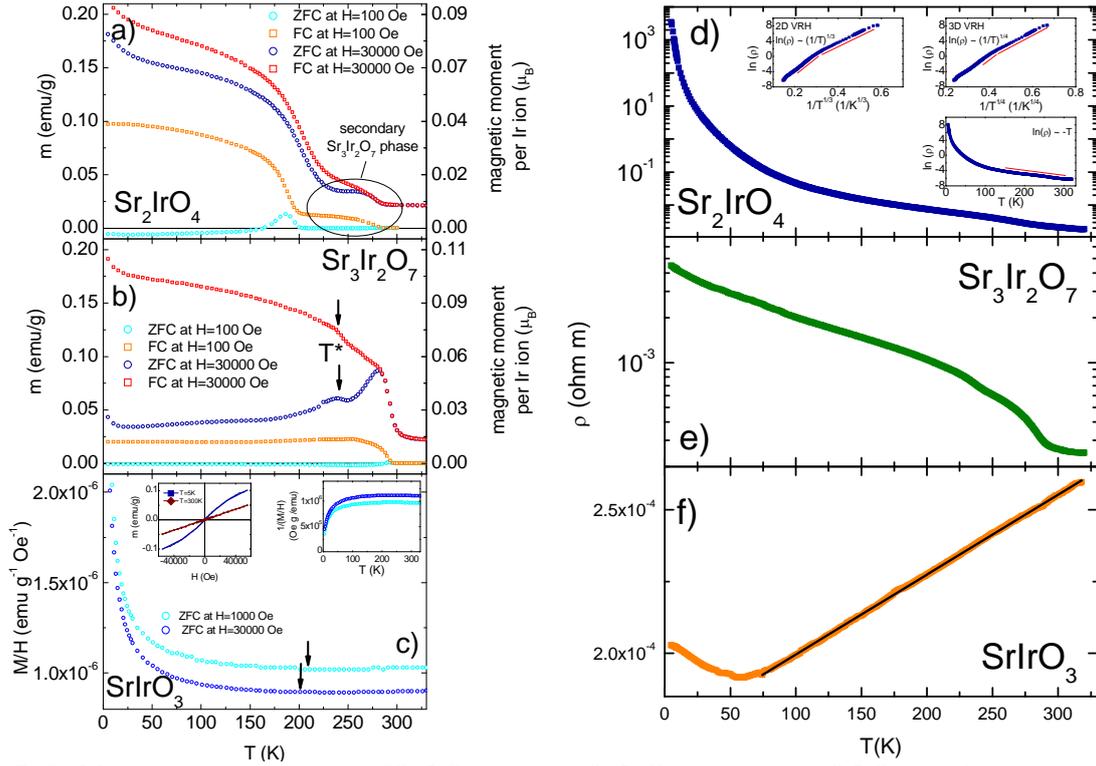

**Figure 3: Left:** Magnetic characterization of $Sr_2IrO_4$ (panel a), $Sr_3Ir_2O_7$ (panel b) and $SrIrO_3$ (panel c) polycrystals. For the first two compounds the magnetic moment as a function of temperature measured at different fields is shown, while for the last one the magnetic susceptibility M/H (main panel c) and inverse magnetic susceptibility M/H (right inset of panel c) are plotted. In panel b), the arrows indicate the magnetic feature at T*. In the main panel c), the arrows indicate an inflection point, dividing the decreasing Curie-Weiss M/H temperature dependence at low temperature from the slightly increasing M/H temperature dependence at high temperature. The magnetic moment as a function of magnetic field, measured at different temperatures is also shown in the left inset of panel c).

**Right:** Transport properties of $Sr_2IrO_4$ (panel d), $Sr_3Ir_2O_7$ (panel e) and $SrIrO_3$ (panel f) polycrystals. In panel d) the resistivity of $Sr_2IrO_4$ is plotted versus temperature in logarithmic scale and in the three insets the plots evidencing possible 2D VHR, 3D VHR and $\ln(\rho) \propto -T$ trends are shown. Red straight lines are guides to the eye to identify the relevant regimes. In panel b) the resistivity of $Sr_3Ir_2O_7$ is plotted versus temperature in logarithmic scale. In panel f) the resistivity versus temperature and power law fit $\rho = \rho_0 + \rho_1 T^\alpha$ (continuous black line) of $SrIrO_3$ are plotted in linear scale.

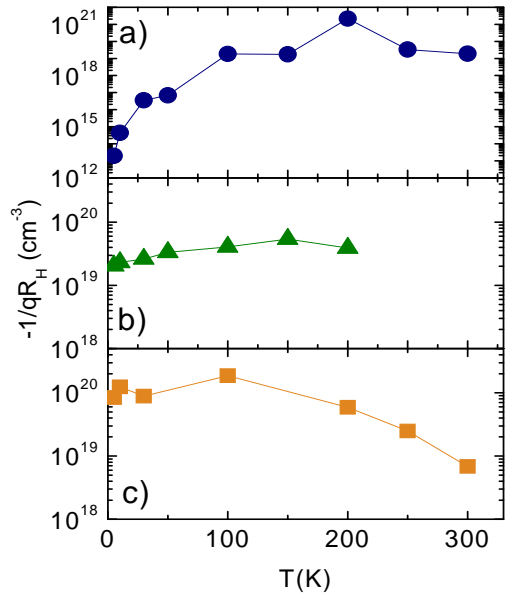

**Figure 4:** Temperature dependence of the inverse Hall resistance of the a) $Sr_2IrO_4$, b) $Sr_3Ir_2O_7$ and c) $SrIrO_3$ polycrystals.

### 3.3 Thermoelectric characterization

Very limited literature on Seebeck effect of $Sr_2IrO_4$, $Sr_3Ir_2O_7$ and $SrIrO_3$ exhists. Indeed, only Seebeck curves measured in $Sr_2IrO_4$ in zero magnetic field can be found [17,26,37,38], showing a positive Seebeck coefficient with a broad maximum around 150K, whose temperature behavior is typical of carrier depletion with decreasing temperature. Yet, thermoelectric characterization reveals very interesting information, as detailed in the following. Before the presentation of Seebeck data, we point out that all transport measurements probe a voltage drop along a percolative path (of the least resistive phase), regardless the driving force is the applied current or the applied thermal gradient. Hence we can assume that if resistivity, Hall resistance and current-voltage data represent on the whole the dominant behavior of the correct phase for each sample, this is also true for thermoelectric data, regardless non negligible contributions of secondary phases are seen in bulk-sensitive X-rays and magnetization measurements. In figure 5a), the Seebeck coefficient curves measured in our $Sr_2IrO_4$ polycrystal in zero and 70000 Oe magnetic fields are shown, well consistent with existing literature data. S is positive in the whole temperature range and its peak value is $\approx 160 \mu V/K$. The non monotonic behavior is explained in terms of competition between two mechanisms, the former dominating below 150K and the latter dominating above 150K, namely: (i) decreasing S with decreasing temperature, related to the vanishing entropy of the charge carriers while temperature is approaching T=0 and (ii) increasing S with decreasing temperature, associated to decrease of thermally activated charge carriers with decreasing temperature (compare figure 4a)). Indeed, regarding the latter effect, for non-degenerate semiconductors in the diffusive regime the inverse dependence of S on carrier density n holds, $S \propto -\ln(n)$ [39]. Hence the Seebeck curves can be explained by the diffusive mechanism alone, keeping into account decrease of carrier density with decreasing temperature observed by Hall effect, without invoking any drag mechanism. The negligible magnetic field dependence confirms the interpretation in terms of diffusive mechanism. The absence of drag terms is consistent with the low density strongly localized character of charge carriers in this compound, as any drag mechanism originates from the coupling of the charge carrier system with a system of boson excitations and hence requires the charge carrier system to be sufficiently populated and capable of exchanging moment with phonons. Noteworthy, S is positive, while the Hall coefficient $R_H$ is negative in this $Sr_2IrO_4$ compound. In a two band description (for $Sr_2IrO_4$ in the simplest picture, a lower Hubbard band of $J_{eff}=1/2$ holes and an upper Hubbard band of $J_{eff}=1/2$ electrons) this situation may occur, because, roughly speaking, holes (h) and electron (e) bands contributions are weighed by $\mu^2 n$ in the expression for $R_H$ [40], which assumes the sign of $(\mu_h^2 n_h - \mu_e^2 n_e)$, while they are mostly weighed by $\mu$ in the expression for S [41], which assumes the sign of $(\mu_h n_h S_h - \mu_e n_e S_e)$. If we consider the temperature regimes where VRH has been identified in resistivity data (namely 5K-40K and 40K-120K, as shown in the insets of figure 3d)), we can fit the VRH Seebeck according to Zvyagin's relationship $S \propto T^{\frac{D-1}{D+1}}$, i.e. $S \propto T^{\frac{1}{3}}$ for 2D VRH and $S \propto T^{\frac{1}{2}}$ for 3D VRH [42]. The fits work well, as seen by the linear trends indicated in the insets of figure 5a), confirming the validity of the VRH regimes identified from the resistivity data. However no appreciable difference in the quality of the two fits is noted, thus not allowing to discriminate between 2D VRH or 3D VRH.

A completely different scenario must be considered for the thermoelectric response of the $Sr_3Ir_2O_7$ compound, shown in figure 5b). First of all the magnitude of S is much smaller than that of $Sr_2IrO_4$, below 10 $\mu V/K$ in the whole temperature range, consistent with the larger carrier concentration. More interestingly, S presents a non monotonic behavior with multiple sign changes, which cannot be explained by the diffusive mechanism alone, even in a multiband description, considering that other electrical properties such as resistivity and charge density exhibit a pretty regular temperature dependence. Such S behavior can be rather explained assuming the contributions of different drag terms, possibly phonon drag and magnon drag, as well as a multiband character of the material. Indeed, drag contributions to Seebeck curves are due to momentum exchange between the charge carrier system and a system of boson excitations and they are characterized by a broad maxium in S occurring at the temperatures where the dominant mechanism by which boson excitations thermalize is the electron-boson scattering, as compared to boson-boson occurring at higher temperatures or boson-defect scattering occurring at lower temperatures. In general, such maximum is expected at temperatures around $\Theta/5$ - $\Theta/10$, where $\Theta$ is the characteristic temperature of the boson excitation spectrum, namely the Debye temperature $\Theta_D$ if the bosons are phonons or the magnon mode energy $\Theta_M$ if the bosons are magnons. The width and position of the maximum may be affected in case of simultaneous presence of phonons and magnons, interacting with each other. In our case, the dependence of S on the magnetic field, seen as a branch-off between S curves measured in zero and 70000 Oe magnetic fields in figure 5b) and more in detail as a parabolic field dependence in the inset of figure 5b), should originate form the magnon drag term. In the case of a generic compound exhibiting AFM exchange coupling, an applied magnetic field would increase spin fluctuations, thus increasing the magnon density, whereas in the case of a generic FM exchange coupling, spin fluctuations would be damped by an applied magnetic field, thus decreasing the magnon density [43]. As a consequence, the magnetic field should increase the magnitude of the magnon drag Seebeck, $S_{MD}$, in AFM compounds, whereas it should decrease the magnitude of $S_{MD}$ in FM compounds [44]. In the plots of figure 5b), it is seen that in the range of positive S, the magnetic field increases the magnitude of S, while in the range of negative S, the magnetic field decreases the magnitude of S. $Sr_3Ir_2O_7$ is a canted AFM, hence it is plausible to expect that the magnon drag Seebeck is enhanced in magnitude by the applied field. Thereby we suggest that the magnon drag Seebeck provides a broadly peaked *positive* contribution below 200K, actually enhanced by magnetic field, in competition with a *negative* phonon drag contribution peaked at 35K and with a diffusive contribution. In this picture, assuming that hole and electron mobilities are not dramatically different [45], the positive sign of the magnon drag and negative sign of the phonon drag indicate that the magnons preferentially couple with $J_{eff}=1/2$ holes, while the phonons mainly couple with $J_{eff}=1/2$ electrons at the respective temperatures where each boson system exchanges moment mainly with charge carriers. Regarding the negative phonon drag term, it has been demonstrated that indeed phonons couple uniquely with the electron band in $Sr_2IrO_4$ and not with the hole band [46]. This is likely to occur also in $Sr_3Ir_2O_7$, where the significant role played by the electron-phonon coupling has been demonstrated as well [28,47]. Moreover, we note that the temperatures of the negative phonon drag peak and positive magnon drag peak are ≈35K and 100K-150K, consistent with the $\Theta/10$ rule of thumb, being in iridates the Debye temperatures $\Theta_D$ around 300K-350K and the typical energies of magnons around 0.15eV-0.25eV [48].

The Seebeck coefficient of the $SrIrO_3$ compound shown in figure 5c) is in the range of few µV/K in magnitude and presents a qualitatively similar non monotonic behavior, which cannot be explained by the diffusive mechanism common in metals, but, as in the case of $Sr_3Ir_2O_7$, can be rather explained assuming the contributions of phonon and magnon drag terms in a multiband scenario. The magnon drag term is responsible for the field dependence of S, also shown in the inset of figure 5c), which is again parabolic. Also for $SrIrO_3$, the non monotonic behavior could be due to the fact that both electrons and holes contribute to transport, and the system of magnons couples preferentially with one type of carriers, while the system of phonons couples with the other type. In analogy with the case of $Sr_3Ir_2O_7$, we may assign preferential coupling of magnon and phonons to holes and electrons, even if in the case of $SrIrO_3$ it is not clear if magnetic interactions are of FM or AFM type. Indeed, the Curie temperature obtained by fitting the our magnetic susceptibility with a modified Curie-Weiss law is nearly zero, which indicates that the magnetic interaction between the two electronic spins on the nearest-neighbor Ir sites is at the crossover between FM-type and AFM-type. We point out that impurity phases cannot be responsible for the qualitative similarity between Seebeck effects in $Sr_3Ir_2O_7$ and $SrIrO_3$ [49]. On one hand, the similarity between phonon drag peaks is not surprising, given the common structural features (it would be likely present also in the $Sr_2IrO_4$ compound, if the carrier density were large enough for drag mechanism to occur). On the other hand, also magnon drag terms are qualitatively similar. On a quantitative basis, any direct comparison between Seebeck magnon drag terms in $SrIrO_3$ and $Sr_3Ir_2O_7$ cannot be made, as the superimposed diffusive contribution to the Seebeck effect is unknown, however it appears that both the peaks attributed to phonon and magnon drag, as well as the variation of the Seebeck coefficient with applied field, are around 3 times larger in $Sr_3Ir_2O_7$ than in $SrIrO_3$. Considering the paramagnetic nature of $SrIrO_3$ versus canted antiferromagnetic nature of $Sr_3Ir_2O_7$, we indeed expect weaker superexchange interactions in the former, and consequently smaller magnon mode energies and magnon density. However, non negligible exchange interactions, and thus non negligible magnon density with respect to a conventional paramagnet, must be present in $SrIrO_3$, as confirmed by the S-shaped m(H) behavior (see left inset in figure 3c)). As the Seebeck magnon drag can be roughly factorized as the product of magnon density and carrier-magnon coupling [44], we gather that in spite of the expected smaller magnon density in $SrIrO_3$, the carrier-magnon coupling in this compound is likely as much strong as in the $Sr_3Ir_2O_7$ compound. Indeed the non Fermi liquid resistivity behavior of $SrIrO_3$ (see figure 3f)) is a signature of the strength of such coupling in this compound. More in general we conjecture that possibly the strong coupling between charge carriers and spin fluctuations is a feature of all Ruddlesden-Popper iridates and may be present also in the $Sr_2IrO_4$ compound, although the exponentially small carrier density in undoped $Sr_2IrO_4$ does not allow to observe any magnon drag Seebeck effect. Possibly any such magnon drag contribution could be observed if the carrier density was increased enough by doping. Noteworthy, the strong coupling between charge carriers and spin fluctuations giving rise to a sizeable magnon drag thermopower has been also observed in the parent compounds of oxypnictide superconductors, where it is believed to play the key role in the pairing mechanism of superconductivity [44]. Also in the case of iridates, such strong coupling is a promising prerequisite for the occurrence of unconventional superconductivity upon doping, indeed predicted theoretically [50] and long sought after in doped $Sr_2IrO_4$.

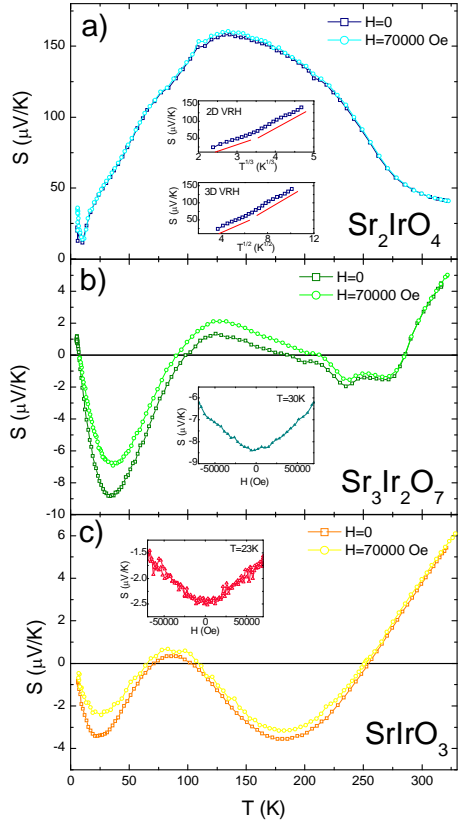

**Figure 5:** Seebeck coefficient curves versus temperature of the a) $Sr_2IrO_4$, b) $Sr_3Ir_2O_7$ and c) $SrIrO_3$ polycrystals, measured in zero and 70000 Oe magnetic fields. Insets of panel a): plots evidencing possible 2D and 3D VHR in the temperature ranges 13K-40K and 40K-100K, also identified in the resistivity curve (see insets of figure 3a)). Red straight lines are guides to the eye to identify the relevant regimes. Insets of panels b) and c): field dependence of S measured at the temperatures where the difference between zero field and in-field S(T) curves is maximum, namely 30K for $Sr_3Ir_2O_7$ and 23K for $SrIrO_3$.

### 3.4 Thermal transport characterization

Figure 6 shows the thermal conductivity κ curves of all the three $Sr_2IrO_4$, $Sr_3Ir_2O_7$ and $SrIrO_3$ samples. The shapes and values of the κ(T) curves are strongly influenced by phonon scattering at grain boundaries, as typical of polycrystalline samples, however the room temperature κ values are larger than those measured in other polycrystals [17], indicating reduced porosity and grain boundary effects in our samples. The room temperature κ values increase with increasing series index, from 2 W/Km in $Sr_2IrO_4$, to 3 W/Km in $Sr_3Ir_2O_7$, to 4 W/Km in $SrIrO_3$. However, this trend should not be related with carrier density, given that κ is totally determined by the lattice contribution for all the samples. Indeed, using the Wiedmann-Franz law, the electronic contribution to κ is estimated to be 500 times, 100 times and 100 times smaller than the measured value at room temperature for $Sr_2IrO_4$, $Sr_3Ir_2O_7$ and $SrIrO_3$ respectively. We tentatively relate the trend of κ with the dimensional character of the lattice, from 2D to 3D with increasing series index. However, considering that the porosity of the microstructure strongly affects the thermal conductivity value, the trend of κ may be also influenced by extrinsic causes. The thermal conductivity is virtually field independent (in-field data not shown), consistent with its phonon nature.

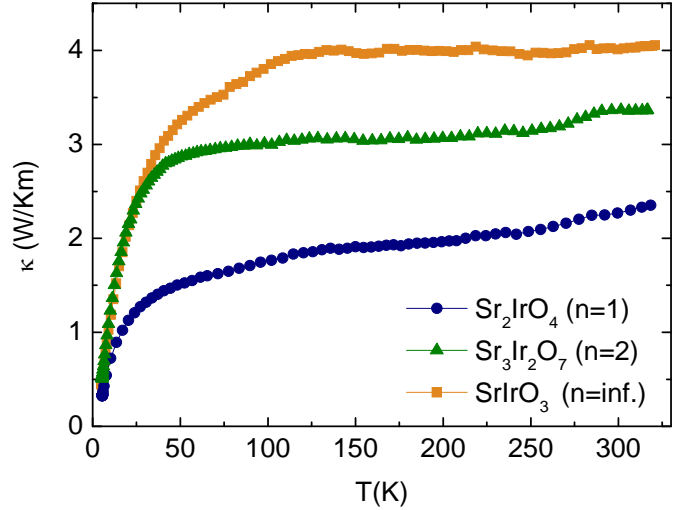

**Figure 6:** Thermal conductivity curves versus temperature of the $Sr_2IrO_4$ (n=1), $Sr_3Ir_2O_7$ (n=2) and $SrIrO_3$ (n=∞) polycrystals.

## 4. Conclusions

We carry out full characterization of structural, magnetic, electric and thermoelectric properties of Ruddlesden-Popper series iridates $Sr_{n+1}Ir_nO_{3n+1}$ (n=1, 2 and=∞) in order to extract information about transport mechanisms across the members of the series, where a metal-insulator transition driven by bandwidth occurs, and about the coupling between electronic and magnetic degrees of freedom.

High pressure (50 kbar) synthesis of polycrystals yields the correct phase, with little amounts of secondary phases. Magnetic behavior is in agreement with existing literature, confirming that $Sr_2IrO_4$ and $Sr_3Ir_2O_7$ are canted antiferromagnets, while $SrIrO_3$ exhibits enhances paramagnetism. The resistivity varies by six orders of magnitude from room temperature to low temperature in the strongly insulating $Sr_2IrO_4$ and is described by the variable range hopping mechanism in a wide temperature regime. The resistivity of $Sr_3Ir_2O_7$ varies in a smaller range within the same temperature window, indicating the proximity to the metal-insulator transition, while $SrIrO_3$ lies at the other side of the metal-insulator transition, exhibiting metallic non Fermi liquid behavior related to coupling with spin fluctuations. For all the compounds, Hall effect data indicate that electrons are the dominant charge carriers, and their density increases with increasing index of the series, being strongly temperature dependent in the insulating $Sr_2IrO_4$. Seebeck coefficient curves reveal intriguing behaviors. While $Sr_2IrO_4$ exhibits a positive S described by the conventional diffusive mechanism, $Sr_3Ir_2O_7$ and $SrIrO_3$ Seebeck curves presents a non monotonic behavior with multiple sign changes, determined by phonon drag and magnon drag contributions, combined with the presence of multiple bands participating in transport. In particular, in $Sr_3Ir_2O_7$ magnons preferentially couple with holes, while the phonons mainly couple with electrons. This new result on thermoelectric response reveals the strong coupling of charge carriers with phonons and spin fluctuations, as well as the relevance of multiband description in these compounds. We believe that the strong coupling of charge carriers with spin fluctuations could be play a key role in the non conventional superconducting pairing mechanism predicted to occur in doped iridates.

# Supplementary data

**Thermoelectric behavior of Ruddlesden-Popper series iridates**

I. Pallecchi [1], M. T. Buscaglia [2], V. Buscaglia [2], E. Gilioli [3], G. Lamura [1], F. Telesio [1], M. R. Cimberle [1], D. Marré [1]

[1] CNR-SPIN and Dipartimento di Fisica, Via Dodecaneso 33, 16146 Genoa, Italy
[2] CNR-IENI, Via De Marini 6, 16149 Genoa, Italy
[3] CNR-IMEM, Area delle Scienze, 43124 Parma, Italy

## 1. Magnetic characterization

The magnetic data measured on our $Sr_2IrO_4$, $Sr_3Ir_2O_7$ and $SrIrO_3$ polycrystals, displayed in figure S1. In the left-hand panels a), c) and e) we show magnetic moment curves as a function of temperature $m(T)$, measured at different fields, in zero-field-cooled (ZFC) and field-cooled (FC) conditions. In the right-hand panels b), d) and f), we show the $m(H)$ loops measured at different temperatures.

In the top panels a) and b), the $m(T)$ and $m(H)$ curves of the $Sr_2IrO_4$ $n=1$ member of the series are consistent with the expected weak ferromagnetism associated to canted AFM [1,2] at temperatures below a double transition at $\approx 190K$ and $\approx 275K$ in a field H=100 Oe (transition temperatures are defined at the derivative peaks of the FC curve). In the $m(T)$ curves measured at H=30000 Oe, the transitions are shifted at slightly higher temperatures $\approx 200K$ and $\approx 280K$. Whereas the lower transition certainly belongs to the dominant $Sr_2IrO_4$ phase, the upper transition is likely associated to minor presence of $Sr_3Ir_2O_7$ phase, detected also in X-rays diffraction patterns and estimated be around 5% in content. At both high ($H$=30000 Oe) and low ($H$=100 Oe) field, the ZFC and FC $m(T)$ curves depart below the higher transition temperature. The FC curves are similar in shape at both fields. On the contrary, the ZFC curves differ significantly, as the one measured at $H$=100 Oe exhibits a peak in correspondence of the lower transition, as expected for sheer AFM transitions, and saturates to a small negative value at low temperature related to core diamagnetism [3], while the ZFC curve measured at $H$=30000 Oe increases weakly and monotonically with decreasing temperature, possibly due to the effect of the larger field on the canting angle. Above the upper transition, the curves are described by a modified Curie Weiss law $\chi = \chi_0 + C/(T-\theta)$, but the particular fitting values are likely representative not only of the main $Sr_2IrO_4$ phase ($T_N\approx 200K$), but also of the secondary $Sr_3Ir_2O_7$ phase ($T_N\approx 280K$). The $m(H)$ loops shown in figure 1b) exhibit clear hysteresis with coercive fields $H_c$ that, as usual, increase with decreasing temperature, namely $H_c\approx 3000Oe$ at $T$=5K and $H_c\approx 700Oe$ at $T$=150K. The observed rapid magnetic saturation is consistent with literature on both polycrystals [3,4] and single crystals [2]. The saturated magnetic moment, as widely observed in literature [3,2], is a small fraction of the value predicted for the $J_{eff}=1/2$ state, which is a consequence of AFM interactions and structural details [5]. No metamagnetic transition is detected up to $H$=55000 Oe, as in other polycrystalline samples [4].

The middle panels of figure S1 report the magnetic characterization of the $Sr_3Ir_2O_7$ $n=2$ member of the series. On the left hand panel, the $m(T)$ curves measured in ZFC and FC conditions in different fields closely resemble the curves measured in single crystals [6], exhibiting a transition to AFM canted ordering around 290K and an additional feature around $T^*\approx 240K$. We note that our data bear no trace of the exotic magnetic behavior with magnetic reversal below 20K under field cooled conditions seen in certain single crystals [7], which on the other hand is observed neither in polycrystals [8], nor in other single crystals [9]. Most features of $m(T)$ curves of $Sr_3Ir_2O_7$ are similar to the case of the $Sr_2IrO_4$ compound, namely the transition temperature is slightly shifted from 290K to 295K with increasing field from 100 Oe to 30000 Oe, the ZFC curves measured at either high ($H$=30000 Oe) or low ($H$=100 Oe) field differ significantly, with the latter saturating to a constant negative value, the ZFC and FC curves depart below the AFM ordering transition. However, in this compound the separation between ZFC and FC curves at low temperature is larger as compared to $Sr_2IrO_4$, indicating a stronger AFM interaction between adjacent Ir magnetic moments. Consistently, the recent finding of a huge magnon gap of 92 meV in $Sr_3Ir_2O_7$ [10], resulting from enhanced pseudodipolar interactions (energy scales for magnetic interaction surpassing those related to charge degrees of freedom) is in sharp contrast with the magnetic spin-wave spectrum of $Sr_2IrO_4$ showing no observable spin-wave gap [11]. Beside the different strength magnetic interactions themselves in the two compounds, also the role of charge carriers, expected to be more relevant in the crossover $Sr_3Ir_2O_7$ compound, is crucial in this context. Indeed, it has been pointed out that the energy scale of the magnon gap indicates that the AFM transition at ≈ 280K is not driven by thermal fluctuations of magnetic moments, but rather by thermally activated charge carriers that form polarons [10]. Incidentally, this is a clue of the strong coupling between charge carriers and spin fluctuations, which is one of the main results of this work, extracted from the thermoelectric data shown in the main paper. Above the upper transition temperature, the *m(T)* curves do not obey the modified Curie Weiss law, possibly as a consequence of a magnetic phase that persists above 600K [12]. The additional feature observed in *m(T)* curves at the characteristic temperature $T^*$≈240K is likely closely related to the anomalous behavior observed in our *m(H)* curves shown in figure S1d). In particular, the hysteresis loops measured close to $T^*$ are not symmetric with respect to the $H$=0 axis, they do not close and they exhibit maximum coercive field. The effect of not closing loops is well evident in the 250K loop and gradually vanishes with decreasing temperature, disappearing completely below 30K, where well behaved hysteresis loops are observed. In addition the hysteresis decreases with decreasing temperature, oppositely to the usual behavior, with coercive fields $H_c$≈10000Oe at 250K, ≈7500 Oe at 120K and 30K and ≈3000Oe at 5K. This phenomenology, also observed in Y [13] and Sm [14] iridates, is compatible with an exchange bias effect occurring around $T^*$, also verified by inspecting the shift of the loops measured after field-cooling at -55000 Oe and +55000 Oe. As the exchange bias implies the presence of interfaces between ferromagnetic and antiferromagnetic orderings, we suggest that such interfaces may form as a consequence of a magnetostrictive distortion manifesting just around $T^*$, which induces a uniaxial distortion of the Ir lattice and alters the canted AFM ordering of certain Ir sites in presence of applied magnetic field. The presence of a magnetostrictive distortion at $T^*$ has been already suggested in [7], to account for the observed thermomagnetic irreversibility of the system cooled in different magnetic fields after being heated above $T^*$. In the loop measured at 250K a broad metamagnetic transition is seen at fields $H$≈25000 Oe, likely a manifestation of the exchange bias like behavior around $T^*$. The *m(H)* loops in figure S1d) measured at low temperature present lenticular shape with no sign of saturation, which confirms the strength of AFM interactions, in agreement with literature on polycrystals [8] and some single crystals [9], while in other single crystals rapid saturation has been observed [7]. Siumilarly to the $Sr_2IrO_4$ phase, the extracted ordered magnetic moments per Ir atom turn out to be a small fraction of the value predicted for the $J_{eff}$=1/2 state, consistent with literature [7,8].

The lower panels of figure S1 present the strikingly different behavior of the $SrIrO_3$ end member of the series. The temperature dependence of the magnetic susceptibility *M/H* curves in figure S1e) and inverse magnetic susceptibility in the inset indicate paramagnetic behavior well described by a modified Curie-Weiss law below ~200K (at 200K *M/H* curves present an inflection point, similarly observed at ~170K in ref. [15]), with an effective magnetic moment per Ir atom $\mu_{eff}$=0.0025$\mu_B$, much smaller than expected, possibly as a consequence of the hybridization between Iridium *5d* and oxygen *2p* orbitals, and a Curie temperature ≈1K, close to zero. Such small Curie temperature suggests that the magnetic interaction between the two electronic spins on the nearest-neighbor Ir sites lies at the crossover between FM-type and AFM-type. In figure S1f), the S-shaped *m(H)* a low temperature indicate exchange enhanced paramagnetism [15]. Our magnetic data on SrIrO$_3$ are fully consistent with literature [15].

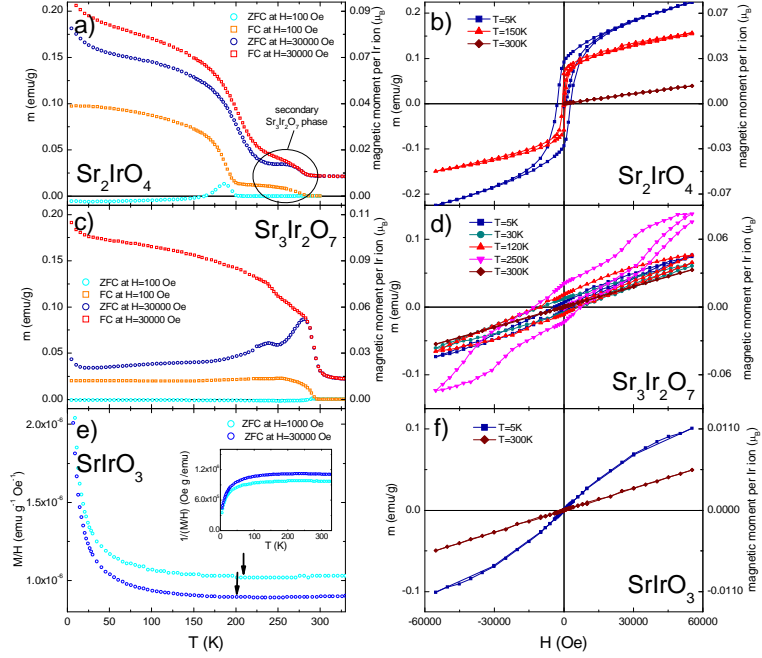

**Figure S1:** Magnetic characterization of Sr$_2$IrO$_4$ ($n$=1) (panels a and b), Sr$_3$Ir$_2$O$_7$ ($n$=2) (panels c and d) and SrIrO$_3$ ($n$=∞) (panels e and f) polycrystals. The left-hand panels show the behavior of the magnetic moment as a function of temperature, measured at different fields, while the right-hand panels show the behavior of the magnetic moment as a function of magnetic field, measured at different temperatures. In the case of the paramagnetic SrIrO$_3$ compound, the magnetic susceptibility $M/H$ (panel e) and inverse magnetic susceptibility $M/H$ (inset of panel e) are plotted rather than the magnetic moment. In the same panel, the arrows indicate an inflection point, dividing the decreasing Curie-Weiss $M/H$ temperature dependence at low temperature from the slightly increasing $M/H$ temperature dependence at high temperature.

## 2. Complete magneto-transport characterization

We now present transport and magneto-transport properties, showing additional magnetoresistance curves and current-voltage characteristics with respect to the main paper.

In figure S2a), the resistivity of the lowest index ($n$=1) member of the series Sr$_2$IrO$_4$ is plotted in logarithmic scale. The insulating behavior and variable range hopping regimes are discussed in the main paper.

In figure S2b), magnetoresistivity $(\rho(H)-\rho(H=0))/\rho(H=0)$ curves measured at selected temperatures are shown. The magnetoresistivity is always smaller than 1% up to 70000 Oe, it is positive of cyclotronic nature at low temperature below 50K, while at higher temperatures ($T \geq 100$K) it is dominated by a negative term. This negative term likely originates from the large spin-orbit scattering, possibly via a anisotropic magnetoresistance (AMR) mechanism [16]. We remark that the field dependence of our magnetoresistivity curves is completely different from the one measured in some single crystals, where it has been attributed to magnetic field-induced lattice distortions [17].

Figure S2c) shows the charge carrier density extracted from the Hall resistance $R_H$ in a single band framework $n = (qR_H)^{-1}$. As discussed in the main paper, the dominant charge carriers are electron-type ($J_{eff}$=1/2 electrons of the upper Hubbard band) and their density decreases significantly with decreasing temperature from ~10$^{19}$ cm$^{-3}$ at room temperature down to ~10$^{13}$ cm$^{-3}$ at the lowest temperature, consistently with the strongly insulating behavior of this compound. To the best of our knowledge, no other Hall effect data of Sr$_2$IrO$_4$ exist in literature. Combining resistivity and charge carrier density data, the Hall mobility $\mu$ is obtained. As shown in the inset of figure S2c), $\mu$ is in the range 0.1-1 $cm^2V^{-1}s^{-1}$, with no clear trend as a function of temperature. Indeed, roughly constant carrier mobility and stongly temperature dependent carrier density correspond to the typical scenario of thermal activation.

The bottom panel of figure S2 displays the current-voltage characteristics plotted as current density $J$ versus electric field $E$, measured at different temperatures. Ohmic (linear) behavior is observed only at 300K, while departure from linearity increases with decreasing temperature and is dramatic at the lowest temperatures below 50K, where negative differential resistivity, not explained by Joule heating effects alone, is observed, as in ref. [2]. In this respect, we point out that all the transport measurements are performed in the ohmic low-$E$ regime.

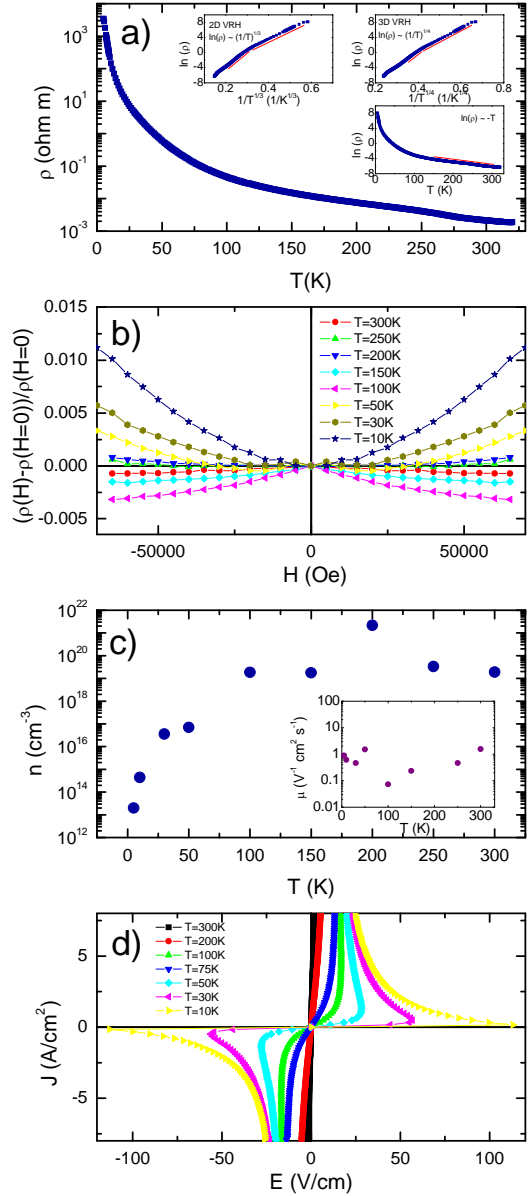

**Figure S2:** Transport properties of the $Sr_2IrO_4$ ($n=1$) polycrystal. a) Resistivity versus temperature in logarithmic scale; in the three insets, the plots evidencing possible 2D VHR, 3D VHR and $ln(\rho) \propto T$ trends are shown. Red straight lines are guides to the eye to identify the relevant regimes. b) Magnetoresistivity $(\rho(H)-\rho(H=0))/\rho(H=0)$ versus field $H$ at selected temperatures. c) Carrier density versus temperature extracted from Hall resistance in a single band approximation; in the inset the Hall mobility is shown. d) Current-voltage characteristics at selected temperatures, plotted as current density $J$ versus electric field $E$.

Figure S3 reports complete transport characterization of $Sr_3Ir_2O_7$, the crossover compound among $n=1$, $n=2$ and $n=\infty$ members of the series, that is the closest to the metal-insulator transition on the insulating side. In figure S3a), the resistivity $\rho$ exhibits semiconducting behavior, but with a much milder temperature dependence as compared to the $Sr_2IrO_4$ compound. Clear changes of slope are observed around 240K and 290K, in correspondence of the magnetic transitions, in agreement with what is widely reported in literature [7,8,9,12,18]. In this respect, although our transport data do not allow to address conclusively the debated issue of the Slater versus Mott character of the insulating state in iridates [19,20], we point out that in the $Sr_2IrO_4$ compound the crossover temperatures between the different VRH regimes are completely unrelated to the magnetic transition temperatures, pointing to Mott character, whereas in the $Sr_3Ir_2O_7$ compound, changes of slope are found in the resistivity curve across the magnetic transitions, below which the AFM state is established. This observation indicates the closer intimacy between electronic and magnetic degrees of freedom in $Sr_3Ir_2O_7$, also confirmed later on in this work by thermoelectric measurements. As discussed in the main paper, this fact, together with the non-linearity of current-voltage curves (see figure S3d)) and the semiconducting behavior, indicates that the overall shape of transport curves must be representative of the correct $Sr_3Ir_2O_7$ main phase, even if a minor contribution of the $SrIrO_3$ impurity phase is certainly present.

The magnetoresistivity shown in figure S3b), of few percent in magnitude, clearly has two contributions, (i) one is positive, proportional to the squared field $H^2$ at low field and slightly tending to saturation at higher fields, hence of cyclotron nature, (ii) the other is negative and appears in the temperature range between 250K and 150K, with maximum intensity around 250K, close to the characteristic magnetic temperature $T^* \approx 240K$ (see figure S1c) and related text). We point out the similarity of our data, with the reports of ref. [12], where the negative magnetoresistivity is maximum close to the characteristic magnetic temperatures and is attributed to the removal of spin disorder scattering by applied field. As for the cyclotron magnetoresistivity term, roughly speaking, the low field $H^2$ magnetoresistivity coefficient is proportional to the carrier mobility, assuming a unknown pre-factor, so that if no dramatic changes in scattering mechanisms occur as a function of temperature, the temperature behavior of the magnetoresistivity coefficient should mirror the temperature behavior of the carrier mobility. The corresponding data are plotted in the inset of figure S3c), showing a mild decrease of magnetoresistivity coefficient with increasing temperature (golden square symbols).

The main panel of figure S3c) shows the charge carrier density extracted from the Hall effect. The negative sign of $R_H$ indicates that the dominant charge carriers are n-type ($J_{eff}=1/2$ electrons of the upper Hubbard band) as in the case of $Sr_2IrO_4$. The charge carrier density is weakly temperature dependent and around few times $10^{19}$ cm$^{-3}$. Differently from the $Sr_2IrO_4$ compound, $n$ does not decrease exponentially with decreasing temperature, indicating the proximity of $Sr_3Ir_2O_7$ to the metal-insulator transition. Combining resistivity and charge carrier density data, the Hall mobility $\mu$ is obtained and plotted in the inset of figure S3c) (purple round symbols). Its value around 1 cm$^2$V$^{-1}$s$^{-1}$ is on the whole slightly larger than that of the $Sr_2IrO_4$ compound and its temperature behavior can be compared with the temperature behavior of the cyclotron magnetoresistance coefficient (golden square symbols). It can be said that in both cases the temperature dependence is weak and the difference in the respective trends may indicate that a single band picture is not fully adequate. A multiband model should be used to describe both magnetoresistivity and $R_H$, however the number of free parameters in this case would be too large to get univocal quantitative information.

In figure S3d), the current-voltage characteristics are ohmic (linear) at high temperature, while mild non-linearity appears only below 30K, consistent with [12].

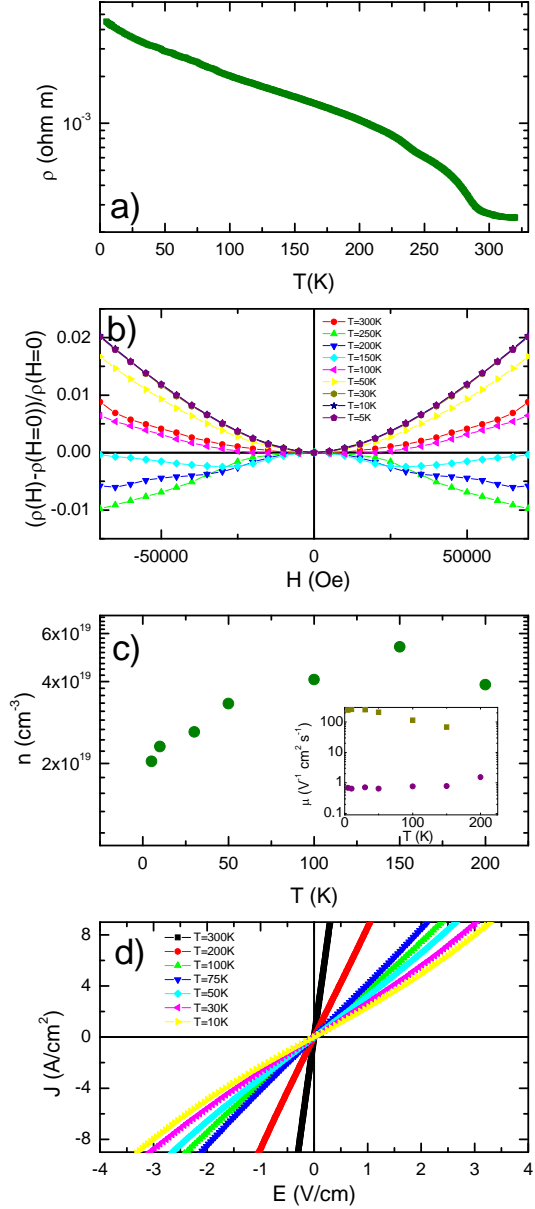

**Figure S3:** Transport properties of the $Sr_3Ir_2O_7$ ($n=2$) polycrystal. a) Resistivity versus temperature in logarithmic scale. b) Magnetoresistivity $(\rho(H)-\rho(H=0))/\rho(H=0)$ versus field $H$ at selected temperatures. c) Carrier density versus temperature extracted from Hall resistance in a single band approximation; in the inset the Hall mobility (purple round symbols) and the coefficient of the low field cyclotron magnetoresistance (golden square symbols) are shown. d) Current density $J$ versus electric field $E$ at selected temperatures.

Finally, the tranport properties of the end member of the series $SrIrO_3$ are presented in figure S4. In the topmost panel, the metallic resistivity is shown, together with a non-Fermi liquid power law fit, discussed in the main paper. Below 50K, the resistivity upturn indicates mild charge localization. Both non-Fermi liquid metallic behavior and low temperature upturn are consistent with literature data on $SrIrO_3$ films [21,22].

The magnetoresistivity, plotted versus the squared field in figure S4b) is positive, proportional to $H^2$ at low field and slightly tending to saturation at higher fields, hence of cyclotron nature. We do not detect any magnetoresistivity component proportional to $H$ which would point to the presence of bands with Dirac dispersion at the Fermi level [23]. The temperature dependence of the low field magnetoresistivity coefficient is shown in the inset of panel c) (golden square symbols), suggesting that the carrier mobility decreases with incresing temperature above 30K, but shows a slight downturn at the lowest temperatures, mirroring the resistivity upturn.

From the negative sign of the Hall resistance $R_H$ it comes out that the dominant charge carriers are n-type (identified also for this compound as $J_{eff}=1/2$ electrons) and the charge carrier density is around $10^{20}$ cm$^{-3}$ at low temperature, one order of magnitude larger than the $Sr_3Ir_2O_7$ compound, as seen in figure S4c), in agreement with literature [22]. The Hall mobility $\mu$ shown in the inset (purple round symbols) is in the range 1-10 cm$^2$V$^{-1}$s$^{-1}$, the largest among these iridates, with no significant temperature dependence. This result, at odds with the decreasing temperature dependence of the magnetoresistivity coefficient, suggests that this material is actually better described by a multiband picture, which is indeed confirmed by the changing sign of the thermoelectric behavior studied in main paper.

For completeness, in figure S4d) we report current-voltage characteristics at two temperatures, both showing ohmic behavior as expected from a metal.

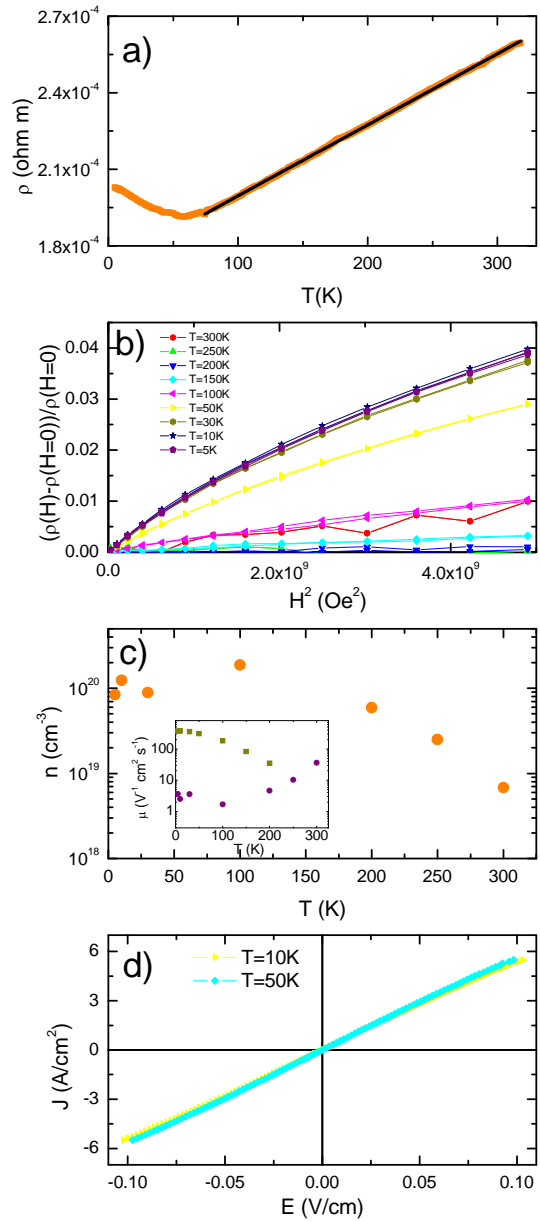

**Figure S4:** Transport properties of the SrIrO$_3$ (n=∞) polycrystal. a) Resistivity versus temperature and power law fit $\rho = \rho_0 + \rho_1 T^\alpha$ (continuous black line). b) Magnetoresistivity (ρ(H)-ρ(H=0))/ρ(H=0) versus squared field H$^2$ at selected temperatures. c) Carrier density versus temperature extracted from Hall resistance in a single band approximation; in the inset the Hall mobility (purple round symbols) and the coefficient of the low field cyclotron

magnetoresistance (golden square symbols) are shown. d) Current density J versus electric field E at two selected temperatures.